\documentclass[twocolumn,showpacs,superscriptaddress,amsmath,aps]{revtex4} %twocolumn,
\usepackage{graphicx}
\usepackage{color}
\usepackage{dcolumn}
\usepackage{mathtools} 
\usepackage{bm}
\usepackage{multirow}
\usepackage{amssymb}
\usepackage{tikz}
\usetikzlibrary{arrows}
\usepackage{subfigure}
\usepackage[colorlinks=true]{hyperref}
\newlength\imagewidth
\newlength\imagescale
\newcommand{\ignore}[1]{}

\begin{document}

\title{The strategy of survival for a competition between normal and anomalous diffusion}
%\title{The Strategy of Survival for a Competition between Multi-agent Diffusion}

\author{Moein Khalighi}
\affiliation{Physics Department, Shahid Beheshti University, G.C., Evin, Tehran 19839, Iran
}
\affiliation{Department of Future Technologies, University of Turku, Finland}

\author{Jamshid Ardalankia}
\affiliation{Department of Financial Management, Shahid Beheshti University, G.C., Evin, Tehran 19839, Iran
}

\author{Abbas Karimi Rizi}
\affiliation{Physics Department, Shahid Beheshti University, G.C., Evin, Tehran 19839, Iran
}

\author{Haleh Ebadi}
\affiliation{Bioinformatics, Institute for Computer Science, Leipzig University,
H\"{a}rtelstrasse 16-18, 04107 Leipzig, Germany
}
\affiliation{Physics Department, Shahid Beheshti University, G.C., Evin, Tehran 19839, Iran
}

\author{Gholamreza Jafari}
\affiliation{Physics Department, Shahid Beheshti University, G.C., Evin, Tehran 19839, Iran
}
\affiliation{Department of Network and Data Science, Central European University, 1051 Budapest, Hungary}

\date{\today}

\begin{abstract}
In this paper, we study the competition of two diffusion processes for achieving the maximum possible diffusion in an area. This competition, however, does not occur in the same circumstance; one of these processes is a normal diffusion with a higher growth rate, and another one is an anomalous diffusion with a lower growth rate. The trivial solution of the proposed model suggests that the winner is the one with the higher growth rate. But, the question is: what characteristics and strategies should the second diffusion include to prolong the survival in such a competition? The studied diffusion equations correspond to the SI model such that the anomalous diffusion has memory described by a fractional order derivative. The strategy promise that anomalous diffusion reaches maximum survival in case of forgetting some parts of the memory. This model can represent some of real phenomena, such as the contest of two companies in a market share, the spreading of two epidemic diseases, the diffusion of two species, or any reaction-diffusion related to real-world competition. 
\end{abstract}


\maketitle

\section{Introduction}\label{sec:intro}
The diffusion processes represent the evolution of many real phenomena, such as epidemic diseases~\cite{Saeedian2017}, gossip spreading~\cite{banerjee2014gossip}, prey-predator species~\cite{bomze1995lotka}, pollution~\cite{gonccalves2013analytical}, and fluid flow~\cite{cussler2009diffusion}. Although there are many approaches in the mathematical view of this context, simple standard mathematical frameworks are inefficient to model some abnormal diffusion processes. The real-world contains eternally competition between the intelligent components of phenomena interacting intellectually in various conditions. Hence, there is still a great demand to advance complex system modeling to interpret such behaviors.

In this paper, we intend to investigate the competition of two normal and anomalous diffusion processes of the SI model. The first diffusion enjoys a higher growth rate, and the other one is an anomalous diffusion including memory. We suggest a master equation which traces the dynamic of the mentioned contest and predicts the future dynamic behavior. It consists of a tunable memory factor that determines the state of ``how much the memory is stimulated, in anomalous diffusion.''
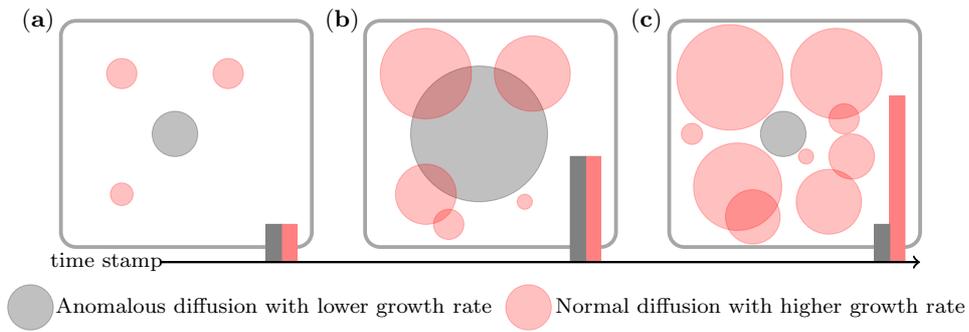
\begin{figure*}[ht!]
\centering
\begin{tikzpicture}

\node at (-.6,2) {(\textbf{a})};
\node at (3.4,2) {(\textbf{b})};
\node at (7.4,2) {(\textbf{c})};
\draw [line width=.5mm,color=gray!70,rounded corners=0.2cm] (-.3,-1) rectangle (3,2) [xshift=4cm] (-.3,-1) rectangle (3,2) [xshift=4cm] (-.3,-1) rectangle (3,2); 

\node at (.3,-1.2	) {\footnotesize{time stamp}};

\filldraw [red!50](2.6,-1.2) rectangle (2.8,-.7);
\filldraw [black!50](2.4,-1.2) rectangle (2.6,-.7);

\filldraw [red!50](6.6,-1.2) rectangle (6.8,.2);
\filldraw [black!50](6.4,-1.2) rectangle (6.6,.2);

\filldraw [red!50](10.6,-1.2) rectangle (10.8,1);
\filldraw [black!50](10.4,-1.2) rectangle (10.6,-.7);

\draw [->,line width=.3mm] (1,-1.2) -- (11,-1.2);

\filldraw [black,nearly transparent](1.2,0.5) circle (.3cm);
\filldraw [red,nearly transparent](1.9,1.3) circle (.2cm);
\filldraw [red,nearly transparent](.5,1.3) circle (.2cm);
\filldraw [red,nearly transparent](.5,-.3) circle (.15cm);

\filldraw [black,nearly transparent](5.2,0.5) circle (.9cm);
\filldraw [red,nearly transparent](5.9,1.3) circle (.5cm);
\filldraw [red,nearly transparent](4.5,1.3) circle (.6cm);
\filldraw [red,nearly transparent](4.5,-.3) circle (.4cm);
\filldraw [red,nearly transparent](4.8,-.7) circle (.2cm);
\filldraw [red,nearly transparent](5.8,-.4) circle (.1cm);

\filldraw [black,nearly transparent](9.2,0.5) circle (.3cm);
\filldraw [red,nearly transparent](9.9,1.3) circle (.6cm);
\filldraw [red,nearly transparent](8.5,1.25) circle (.7cm);
\filldraw [red,nearly transparent](8.6,-.2) circle (.58cm);
\filldraw [red,nearly transparent](8.8,-.6) circle (.36cm);
\filldraw [red,nearly transparent](9.8,-.4) circle (.43cm);
\filldraw [red,nearly transparent](10,.7) circle (.2cm);
\filldraw [red,nearly transparent](8,.5) circle (.14cm);
\filldraw [red,nearly transparent](9.5,0.2) circle (.1cm);
\filldraw [red,nearly transparent](10.1,0.2) circle (.3cm);

\filldraw [black,nearly transparent](-.7,-1.8) circle (.3cm);
\node at (2.5,-1.8) {\footnotesize{Anomalous diffusion with lower growth rate}};
\filldraw [red,nearly transparent](5.85,-1.8) circle (.3cm);
\node at (8.9,-1.8) {\footnotesize{Normal diffusion with higher growth rate}};
%\draw [line width=.5mm,color=gray!70,rounded corners=0.1cm](-.3,-2.9) rectangle (.35,-2.3);
%\node at (2.55,-2.6) {\footnotesize{Capacity of potential customers}};
\end{tikzpicture}
\caption{\textbf{Schematic competition of normal and anomalous diffusion.} \textbf{(a)} At the first stage, the proportion of the anomalous diffusion is supposed to be small and equal to the proportion of the other side of the competition--normal diffusion process with the higher growth rate. \textbf{(b)} The conflict starts when some sharing diffusion areas are emerging, and the growth of one diffusion decays the proportion of another diffusion process. \textbf{(c)} The larger part of the competition establishes an ever-growing behavior so that the anomalous diffusion is likely to inevitable vanishing.}
\label{fig:scheme}
\end{figure*}

The trivial outcome of our proposed model is illustrated in Fig.~\ref{fig:scheme}.
The normal diffusion with a higher growth rate will occupy the more region of the system and maintain its growth.
The counter-side of the rivalry, the one with a lower growth rate, is vulnerable to vanishing. However, by taking into account the memory effects~\cite{Ebadi2016,Saeedian2017,pone} in the anomalous diffusion, it is promising to extend the time interval of maintaining its minimum proportion.

It is worthy to shed light upon possible applications of our proposed model in the industries and lay beyond the reach of theoretical aspects, namely competitive financial interactions~\cite{Iranzo2016,Iori2015}, social marketing events~\cite{Thackeray2008,Ashley2014}, sales promotion which may be applied in a saturated market~\cite{Kaushik2019}, and the new phenomenon so-called \textit{crowd-funding} and financing state-of-the-art technologies~\cite{Kaplan2013}. As well, the proposed idea is not only limited to economics but also extended to other fields of study involving an analogous model. 

In the following, section \ref{sec:def} deals with introducing the master equation with integer order and analyzing its dynamic behavior. In section \ref{sec:fractional}, the differential equation associated with lower growth rate is incorporated into the concept of memory by applying \textit{Caputo} approach~\cite{podlubny1999fractional} to provide the anomalous diffusion. To optimize the memory effects, a strategy will be suggested in section \ref{sec:strategy}, and its quality will be checked in section \ref{sec:heatmap} for the application in business. In section \ref{sec:sensitivity}, the conclusions and future directions are taken.

%%%%%%%%%%%%%%%%%%%%%%%%%%%%%%%%%%%%%%%%%%%%%%%%%%%%%%%%%%%%%%%%%%%%%
\section{Modeling the competition} \label{sec:def}
%%%%%%%%%%%%%%%%%%%%%%%%%%%%%%%%%%%%%%%%%%%%%%%%%%%%%%%%%%%%%%%%%%%%%
Let us denote the normal and anomalous diffusion at time $t$, respectively, by $I_1(t)$ and $I_2(t)$. We consider $S(t) \geq 0$ as a potential shared source at time $t$. We define constant coefficient $\gamma$ referring to the \textit{relative growth rate}, the proportion of the anomalous diffusion in respect to normal diffusion stating on the other side of the competition.

Since the size of the whole system is assumed to be constant, the summation over the amount of the two sides, $I_1(t)$, $I_2(t)$ and the potential capacity $S(t)$ are not independent, so we consider the normalized form satisfying: 
\begin{equation}
1  =  S(t)+I_1(t)+I_2(t)-(I_1(t)\cap{I_2(t)}).
\end{equation}

Each part of the source may distribute to the both diffusion through time. Thus, the growth of $I_1$ and/or $I_2$ leads to the reduction of $S$. Hence, we define the dynamic behavior of the potential capacity $S(t)$ with the following master equation,
\begin{equation}\label{eq:1}
\frac{dS}{dt}  = -( I_1 + \gamma I_2 )S.
\end{equation}

The conversion rate of \(S\) to the two diffusion depends on the growth rate coefficients and the potential capacity. On the other hand, the growth of $I_2/I_1$ should decay $I_1/I_2$, and vice versa. Therefore, one can formulate the dynamics of each diffusion as:
\begin{equation}\label{eq:2}
\frac{dI_1}{dt}  =  (1 - \gamma)I_1 I_2 +  I_1 S, 
\end{equation}
\begin{equation}\label{eq:3}
\frac{dI_2}{dt}  =  (\gamma- 1)I_1 I_2 + \gamma I_2 S,
\end{equation}

By assuming $0<\gamma<1$, the growth rate of diffusion \(I_1\) is higher than diffusion \(I_2\). Under the condition of $\gamma=1$, the two dynamical equations turn into two equal coupled differential equations. In this case, with the same initial values of $I_1$ and $I_2$, the two competitors will grow symmetrically as long as half of the system is occupied.  
\begin{figure*}[ht!]
\centering
\includegraphics[scale=0.5]{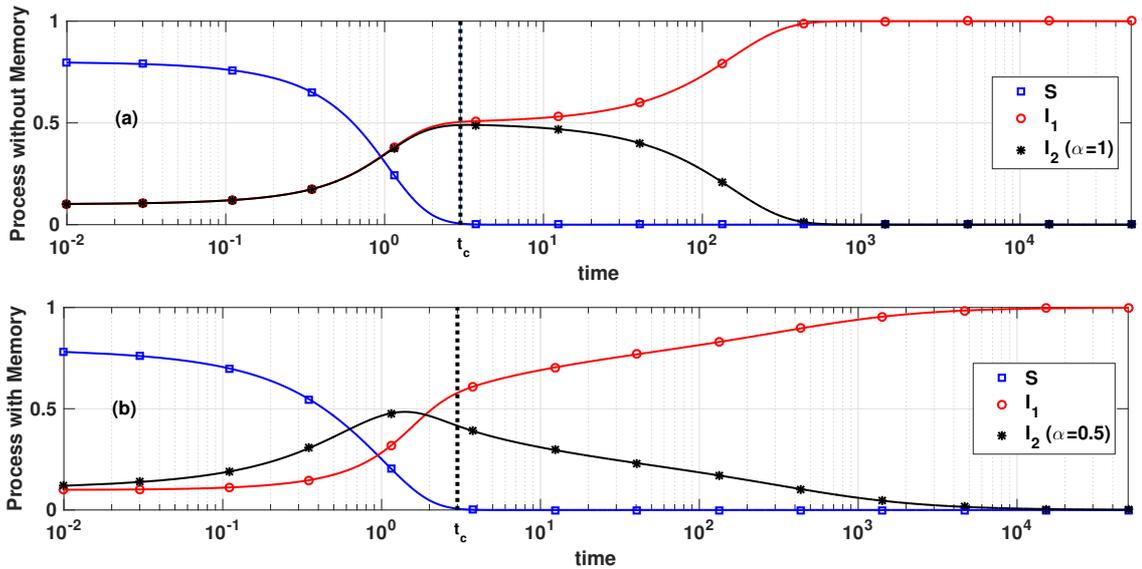}
\caption{
The evolution of $S(t)$, $I_1(t)$ and $I_2(t)$ with the relative growth rate $\gamma=0.995$ with the initial values are $S(0) = 0.8$, $I_1(0) = I_2(0) = 0.1$. \textbf{(a)} The numerical solution of a Markov process based on Eq.\ref{eq:1}, Eq.\ref{eq:2} and Eq.\ref{eq:3}. \textbf{(b)} The numerical solution of a Non-Markov process based on Eq.\ref{eq:11}, Eq.\ref{eq:21} and Eq.\ref{eq:4} with $\alpha = 0.5$.
}
\label{fig:1}
\end{figure*}

In Fig.~\ref{fig:1}, the dynamic of growth and decay of the two diffusions with the same initial value $I_1(0)= I_2(0)=0.1$ and relative growth rate $\gamma=0.995$ show the emerging pattern of the competition to earn a more shared area. $I_2(t)$ reaches a maximum value at critical time $t_c$ where $I_1(t_c)+I_2(t_c) \simeq 1$ and $S(t_c)\simeq 0$. In the case of memory-less, Fig.~\ref{fig:1}(a), the competition between the two sides begins at $t_c$. At this time, the side 1 begins growing faster than side 2 and obtains a bigger region of the system. However, the weaker side, \(I_1\), follows a decreasing trend. Hence, a small difference between the growth rate coefficients of the competitors causes two diverse destinies. Thus, the more powerful the side of the competition will monopolize the system. It shows that the relative growth rate plays a significant role in the success and failure of competitors so that relatively smaller ones have no chance to survive under the competition with bigger rivals.

All the above discussion are based on the defined set of dynamical equations \ref{eq:1} to \ref{eq:3}. The proposed system can be validated by a well-known biological model with a similar concept; In fact, equations \ref{eq:2} and \ref{eq:3} are analogous to Lotka-Volterra competition model~\cite{bomze1995lotka}.
Furthermore, in Sec.~\ref{sec:heatmap}, we will discuss the future states of the temporal contest while the relative growth rate $\gamma$ changes from 0 through 1.
The main question is that which conditions aim the weaker competitor to survive more? 
In the following sections, we will propose a strategy on memory effects to prolong the survival of the weaker competitor. 

%%%%%%%%%%%%%%%%%%%%%%%%%%%%%%%%%%%%%%%%%%%%%%%%%%%%%%%%%%%%%%%%%%%%%
\section{Memory effects} \label{sec:fractional}
%%%%%%%%%%%%%%%%%%%%%%%%%%%%%%%%%%%%%%%%%%%%%%%%%%%%%%%%%%%%%%%%%%%%%
A reaction-diffusion system which includes intelligent elements is affected by memory. However, the proposed model \ref{eq:1}-\ref{eq:3} described by integer order derivatives cannot perfectly describe processes with memory (non-Markovian processes)~\cite{Saeedian2017, podlubny1999fractional}, due to this fact that such derivatives are determined by only a very small neighborhood around each point of time. 

To overcome this shortcoming, we incorporate the concept of \textit{fractional calculus} into the system as a kernel of the differential operator--that is, substituting a fractional order derivative. Indeed, it is shown that fractional derivatives can appropriately represent the effects of power-law memory~\cite{kilbas2006theory,pone,Saeedian2017}. Hence, we consider memory effects only for the evolution of the weaker competitor, $I_2$. As a result, intellectual behaviors that aim to slow down the diffusion decaying can be formulated by applying the memory effects. 

Mathematically, an integral equation with a time-dependent kernel $\kappa(t-t')$~\cite{Saeedian2017,Hassanibesheli2017} enables us to take the effects of previous time steps into account:
\begin{equation}\label{eq:41}
\frac{dI_2}{dt}  =   \int_{t_0}^{t} \kappa(t-t') H dt',
\end{equation}
where
\begin{equation}\label{eq:42}
H=((\gamma-1) I_1(t') I_2(t'))+\gamma I_2(t')S(t'),
\end{equation}
and we set the kernel as:
\begin{equation}\label{eq:43}
\kappa(t-t')=\frac{1}{\Gamma(\alpha-1)(t-t')^{\alpha-2}},
\end{equation}
where $0<\alpha\leqslant 1$ and $\Gamma$ denotes the Gamma function. Different types of fractional differential operators that are suggested by Riemann, Liouville, Grunwald, Letnikov, Sonine, Marchaud, Weyl, Riesz, Caputo, Fabrizio, Atangana, and other scientists~\cite{samko1993fractional, podlubny1999fractional, kilbas2006theory, fabrizio1, atangana2016chaos}. But, in this paper, we consider the Caputo fractional time derivative of order $\alpha$ which can describe physical meanings of real-world phenomena~\cite{podlubny1999fractional}:
\begin{equation}\label{caputo}
{}_{{t_0}}^cD_t^\alpha y(t) = \frac{1}{{\Gamma (\alpha  - 1)}}\int_{{t_0}}^t {\frac{{y'(\tau )d\tau }}{{{{(t - {t_0})}^\alpha }}}} .
\end{equation}
A lower degree of the fractional derivative $\alpha$ indicates a ``stronger" (long-lasting) memory effects of the weaker competitor, $I_2$. Hence, the dynamical equation of $I_2$ will follow a fractional differential while the two other dynamical equations~\ref{eq:1} and~\ref{eq:2} will remain unchanged:
\begin{align}\label{eq:11}
&\frac{dS}{dt}  = - (I_1 + \gamma I_2 )S,\\
\label{eq:21}
&\frac{dI_1}{dt}  =  (1 - \gamma)I_1 I_2 + I_1 S,\\
\label{eq:4}
& {}_{t_0}^cD_t^\alpha I_2(t)  =  (\gamma - 1)I_1 I_2 + \gamma I_2 S.
\end{align}

\begin{figure*}[ht!]
   \centering
    \pgfmathsetlength{\imagewidth}{\linewidth}%
    \pgfmathsetlength{\imagescale}{\imagewidth}%
    \begin{tikzpicture}[x=\imagescale,y=-\imagescale]   
        \node[anchor=north west] at (0,0) {\includegraphics[scale=0.4]{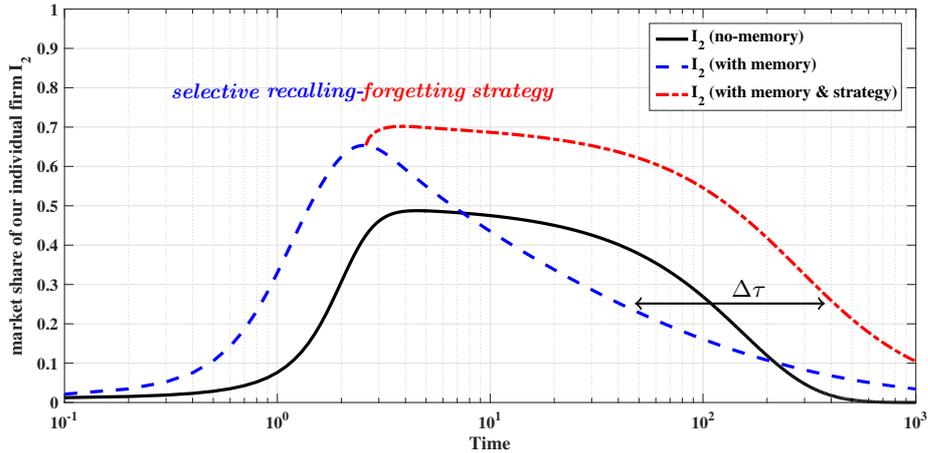}};
       
      \draw [<->][thick](9.5cm,-4.5cm) -- (12cm,-4.5cm);
           % \draw [dotted][thick](2.5cm,-5.23cm) -- (11cm,-5.23cm);
%            \node  at (3.2cm,-5cm) {\textbf{$\xi=0.3$}};
 \node at (11cm,-4.3cm){\textbf{$\Delta\tau$}};
%  \node [text=red] at (6cm,-1.7cm){\textbf{\footnotesize{selective recalling-forgetting strategy}}};
\node [text=blue] at (3.99cm,-1.675cm) {\textbf{\textit{\footnotesize{selective}}}};
\node [text=blue] at (5.3cm,-1.7cm) {\textbf{\textit{\footnotesize{recalling-}}}};
\node [text=red] at (6.6cm,-1.7cm) {\textbf{\textit{\footnotesize{forgetting}}}};
\node [text=red] at (7.9cm,-1.7cm) {\textbf{\textit{\footnotesize{strategy}}}};
%  \draw [->,red,line width=1pt] (6cm,-2.4cm) -- (5.5cm,-1.8cm);
\end{tikzpicture}
\caption{A comparison of the evolution of the anomalous diffusion $I_2(t)$ with the relative growth rate $\gamma=0.995$ and initial value $I_2(0) = 0.01$ for three cases, without memory, with memory, and including memory and strategy. The non-fractional value of $\alpha=1$ guarantees the absence of memory effects in the growth process of $I_2$ (solid black line). The blue dashed line indicates the growth of $I_2(t)$ with the memory factor $\alpha = 0.5$. The red dashed and dotted line corresponds to the growth of \(I_2(t)\) with a new memory which is started at the peak of the memory process with  $\alpha = 0.5$. The interval $\Delta\tau$ denotes the added lifetime for a predefined minimum proportion after launching the strategy.
}
\label{fig:2}
\end{figure*}

    For simplicity, we assume that the memory of Eq.\eqref{eq:4} is constant through time. Thus, by considering $\alpha=0.5$, the emerging competitors start developing with an almost similar rate and an equal potential source converting to two sides by considering the effect of memory, as illustrated in Fig.\ref{fig:1}(b). Interestingly, the influential memory affects the contest before the $t_c$, when the whole source is completely divided into two competitors. It reduces the negative slope of the curve and slows down the loss rate of the weaker side, and hinders the growth of the powerful side. Nevertheless, it is not possible to alter the final destiny of the weaker competitor.  Therefore, after a comparatively longer time, the weaker side inevitably loses its whole system share, and the more powerful side of the competition earns all capacity.

\section{Strategy} \label{sec:strategy}
\textit{Besides remembering, forgetting is a priceless gift of human beings.}\\
We optimize the diffusion behavior by renewing the memory at a particular moment. This strategy may lead the growth curve to the highest level of curves based on different memory stages. 
Initiating the memory from different spots of the functional history timeline of the diffusion and drawing the corresponding curves enables us to compare the growth patterns depending on the memory start point. Such a selective strategy is an approach to remarkably extend the survival time of anomalous diffusion.

Fig.\ref{fig:2} illustrates a comparison of the behavior of the system including memory and strategy (red dashed and dotted line), only memory (blue dashed line), without memory (black solid line), which lead to different growth dynamical curves. 

The black diagram shows the evolution of $I_2(t)$ with the relative growth rate $\gamma=0.995$ with the initial value $I_2(0) = 0.01$ and $\alpha = 1$. The non-integer value of $\alpha$ does not guarantee long-standing survival time, due to the absence of the memory effects in the growth process of $I_2$. 

The blue curve indicates the growth of $I_2(t)$ with a similar relative growth rate and initial values, when the memory is set $\alpha = 0.5$ for the operator $_0^cD_{10^{3}}^\alpha$. In this case, the proportion of the memory-less process lower than the process with memory, however, it achieves a local success after the peak (the advent of the conflict). 

The red curve corresponds to the anomalous diffusion with a new memory starting from the peak of the process with memory. As a result, to extend the survival time of diffusion with a lower growth rate, the anomalous diffusion should continue until the peak point with recalling the past states, then, the process restarts by forgetting past experiences, and a new anomalous diffusion continues the process with considering memory effects from the last peak. To do so, we can determine the fractional differential operator by piecewise functions, $_0^cD_{{t^*}}^\alpha$ and $_{{t^*}}^cD_{{{10}^3}}^\alpha$, where $t^*$ denotes the peak point.

We call this approach ``selective recalling-forgetting strategy" which may indicate some well-known intelligent reactions in the context of Business or other possible aspects. 
Furthermore, despite the maximum value of $I_2$, examining this strategy for two other moments are interesting for advanced complex models; 1.~At the inflection of the curve $S$, when the evolution behaviors are changing. 2.~At the intersection of $I_1$ and $I_2$, when the source is saturated, and both sides of the contest include an equal value.

\section{A Proof of concept} \label{sec:heatmap}

To interpret an application of the main idea, let us assume a business case of study focusing on the competition of two newly founded companies. Hence, we introduce a simple dynamical model to compare the behavior of a multi-agent competing market containing two sides: our individual firm, \(I_2\),  on one side, and the whole market, \(I_1\), except the so-called individual firm, on the other side (see Fig.~\ref{fig:scheme}, and Eqs.~\ref{eq:11}-\ref{eq:4}). By considering whole system as a \textit{market share}, our results will build a bridge connecting a \textit{rivalry of possessing market share} and \textit{fractional calculus}.\\
Therefore, we have analogously discussed:\\
\indent I. the temporal properties of this multi-agent contest;\\
\indent II. the memory effects of one diffusion on the evolution of the whole system;\\
\indent III. by changing the strategy, the extent which anomalous diffusion can sustain in the temporal contest to possess at least a minimum \textit{ad hoc} market share for a longer time;\\

Further discussion is the phase spaces of $\alpha$, $\Delta\tau$, $\gamma$. The notation $\alpha$ is a tunable memory factor that determines the state of how much the memory is stimulated in the weaker firm customers'. Also, $\Delta\tau$ denotes the added lifetime after launching the strategy. $0<\gamma<1$ refers to the relative growth rate of the market share of our individual firm concerning the relative growth rate of the market share of the other side of the competition (the whole market except our individual firm). We have revealed \(t_c\) in Fig.~\ref{fig:1} as a \textit{critical time}, in which the whole potential market is occupied by the competitors and achieving more market share for one firm. It yields to giving up the market share for another firm in the contest. Accordingly, a zero-sum gain~\cite{Krishnamoorthy2010,Hu2010} will emerge. 

As we have theoretically shown, the counter-side market with a higher growth rate will occupy the whole market and maintain their growing market share influenced by advertisements, financial investments~\cite{Krishnamoorthy2010,Hu2010}, hub-connections and united competitors~\cite{Iranzo2016}, so forth. On the other side of the rivalry, our individual firm with a lower growth rate is vulnerable to its market share extinction. Further, by taking into account the memory effects in the weaker firm, it can extend the time interval, $\Delta\tau$, of the minimum market share (Fig. \ref{fig:2}).

To compare the total number of achieved customers of the weaker company, $I_2$, for three different cases--that is, the model without memory ($NMI_2$), with memory ($MI_2$), and with memory and strategy ($SMI_2$), we suggest considering cumulative market share through the time. 
Hence, we denote cumulative function by ``$\int$".
\begin{figure}[ht!]
	\centering
\includegraphics[scale=0.38]{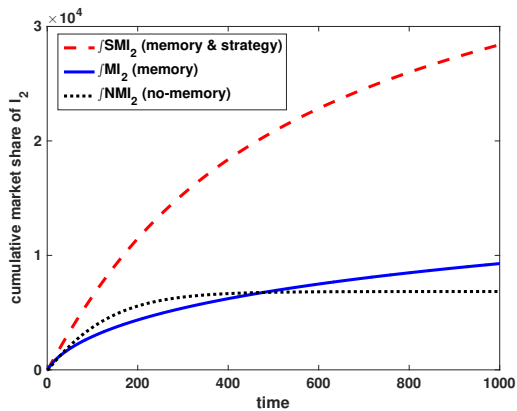}
\caption{
A comparison of cumulative market shares of $I_2$ for three different cases; with memory and strategy, only with memory, and without memory, when $\gamma=0.995$.
}%
\label{fig:3}%
\end{figure}

\begin{figure}[ht!]
\centering
    \pgfmathsetlength{\imagewidth}{\linewidth}%
    \pgfmathsetlength{\imagescale}{\imagewidth}%
    \begin{tikzpicture}[x=\imagescale,y=-\imagescale]   
        \node at (0,0) {\includegraphics[scale=0.45]{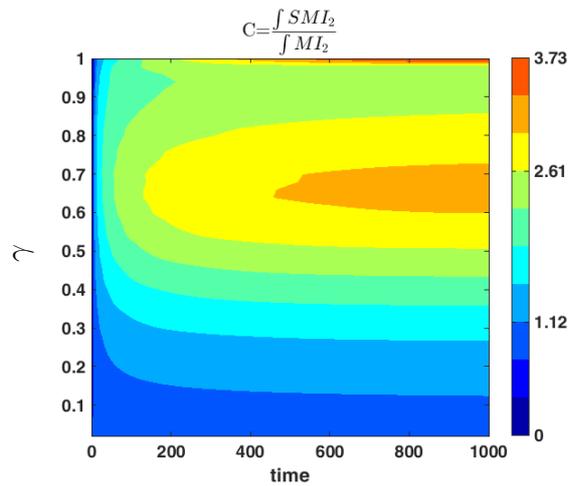}};
 \node [rotate=90] at (-4cm,0cm){\large{\textbf{$\gamma$}}};
\end{tikzpicture}
\caption{Proportions of cumulative market shares of $I_2$, for the system including memory and strategy to the system with memory, in a range of relative growth rates $0<\gamma<1$ through the time-stamp 1000. }%
\label{fig:5}%
\end{figure}

\begin{figure}[ht!]
\centering
    \pgfmathsetlength{\imagewidth}{\linewidth}%
    \pgfmathsetlength{\imagescale}{\imagewidth}%
    \begin{tikzpicture}[x=\imagescale,y=-\imagescale]   
        \node at (0,0) {\includegraphics[scale=0.35]{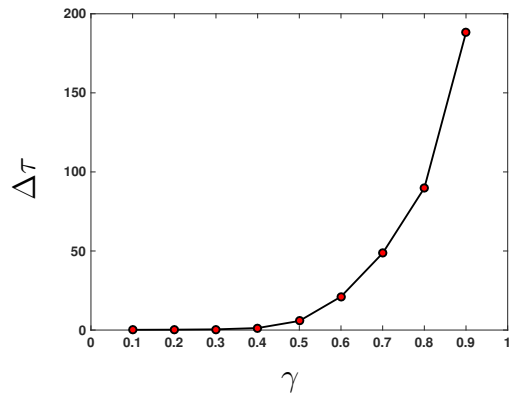}};
 \node at (0cm,-2.7cm){\large{\textbf{$\gamma$}}};
  \node [rotate=90] at (-3.5cm,0cm){\large{\textbf{$\Delta\tau$}}};
\end{tikzpicture}
\caption{Predicting the effect of triggering the new strategy on the lengthening the additional survival time, $\Delta\tau$ (see Fig.~\ref{fig:2}), of the weaker side (our individual firm) for different rates of competitions, $\gamma$.}%
\label{fig:detaT}%
\end{figure}

Fig.~\ref{fig:3} shows that the evolution process involving the strategy (red dashed line) performs better than two other cases, as well the memory influences the system (blue solid line) after around 500. It confirms that, for such a $\gamma$ closing to 1, it is recommended to run the strategy because the impact of using strategy and memory is more than the effects of exclusive memory. Consequently, when the competition between the two firms is too tight (e.g. for $\gamma=0.995$), it is plausible to introduce a selective recalling-forgetting strategy. 

Besides, to clarify the efficiency of the proposed model for various relative growth rates, we provide a heatmap of the proportions of cumulative market share for different competition ranges, $0<\gamma<1$, versus time (Fig.~\ref{fig:5}). 
The notation  ${C} = \frac{{\int {SM{I_2}} }}{{\int {M{I_2}} }}$ indicates a proportion of the cumulative market share of $I_2$ including strategy and memory over the cumulative market share of $I_2$ only with memory. Based on Fig.\ref{fig:5}, for the range of $0.6<\gamma<0.7$ and $\gamma\simeq1$ using a selective recalling-forgetting strategy is highly recommended for surviving.

Considering a predefined minimum market share, Fig.~\ref{fig:detaT} demonstrates the effect of triggering the new strategy on the lengthening the additional survival time ($\Delta\tau$) of the weaker side (our individual firm). When it comes to a lower ratio of relative growth ($\gamma \to 0$), the managers may be reluctant to run the strategy. Because, when $\gamma \to 0$, it results in too small additional survival time ($\Delta\tau \to 0$). However, for larger values of $\gamma$, managers can provide an \textit{trade-off} analysis~\cite{Ardalankia2019} to evaluate the probable profitability.
%%%%%%%%%%%%%%%%%%%%%%%%%%%%%%%%%%%%%%%%%%%%
%%%%%%%%%%%%%%%%%%%%%%%%%%%%%%%%%%%%%%%%%%%%
\section{Discussion} \label{sec:sensitivity}
%%%%%%%%%%%%%%%%%%%%%%%%%%%%%%%%%%%%%%%%%%%%%%%%%%%%%%%%%%%%%%%%%%%%%
The diffusion problems in the real world have always consisted of a competition between various diffusion processes. These competitions occur in varied circumstances; one competitor may have a higher growth rate (or higher diffusion velocity), and the other one surpasses alternative factors. Hence, we have developed a deterministic model of such unequal competitions and studied its dynamic behavior.

Here, a competition model has been proposed in two distinctive processes--without memory effects (normal diffusion) described by integer order differentials, and with memory effects (anomalous diffusion) by non-integer order differentials. We have revealed the impact of memory effects on the competition dynamics and presented a novel strategy by renewing memory effects imposed on the anomalous diffusion.

In the memoryless process, both processes reach a maximum value when the conflict began. After this time, the diffusion processes diverge exponentially so that the more powerful side, even for relative growth rate $\gamma\simeq1$, would dominate the whole system. Thus, the weaker, anomalous diffusion has no chance to survive under the competition with the other rivals on the bigger side. However, there are some factors in real intelligent interactions that moderate such extreme divergence dynamics and we have represented this fact by memory effects.

The proposed model has illustrated that the presence of memory leads to more sustainable dynamics, whereas the lack of memory leads to more energetic dynamics. In this regard, when the process is decaying (or growing), the memory effects have a conservative action on the dynamic. By taking to account such a mechanism, we have prolonged the survival time of the anomalous diffusion.

One application of this strategy makes sense in Business; We maximize the efficiency of an individual weaker venture (relative to the whole market) by recalling the past until the peak point achieved and forgetting the past experiences, and the process is continued with a new memory starting from the last peak. Here, we have suggested that the relative growth rate coefficients can play the role of trade-off effects between value and cost of individual customers~\cite{Ardalankia2019} and it is plausible that the memory~\cite{pone,Saeedian2017,Ebadi2016} represents the characteristics of the value-cost trade-off and provides the customers to satisfy their utility~\cite{Grauwin2009}. 

At the heart of this approach, we emphasize that exploring a new strategy and also other striking actions take time to propagate in society, and this time-lag must be considered~\cite{Banerjee2013}.
Considering scarce resources, two growing economic sectors in a selfish interaction~\cite{Grauwin2009} contribute to a competition of gaining the possible maximum market share and customers. Throughout a certain real-world network of competing agents, in spite of cumulative growth~\cite{Newman2001,Newman2003,Barabsi1999,Barabaacutesi1999}, there may exist some \textit{frictions} and drivers which affect the growth~\cite{pone}. Following this train of thought, there exist internal and external dynamics that create the cost of growth. Accordingly, the states of failure to possess a certain market share, and ever-growing market share, or even a trade-off between further growth or failure in a temporal behavior will emerge. Considering the memory of systems as a decaying factor against sudden alterations~\cite{pone,Hassanibesheli2017}, besides with probable strategies~\cite{Iranzo2016} as a temporal game-changer, in this study, we have applied the memory created by an individual firm--in \textit{statue quo}--in the customers' viewpoint or launching new strategies in the firms as an advantage to compete against the whole market.

To demonstrate the competitors' behavior, some scholars considered restricted areas exposed to overcrowding~\cite{Forgerini2014}. In this context, the systems increasingly grow over time~\cite{pone}. As soon as the accessible region reduces, newer agents may locate in the territory of others, or their territory squeeze. Due to lack of resources--the density of the spatial area around agents--the involving agents are eliminated. This phenomenon will amplify when the space of the contest reduces. Indeed, after a critical time, the systems are vulnerable to some effects against growth, say lack of space in a rivalry and squeezed territories~\cite{Forgerini2014} or the cost of promotion, or agents extinction~\cite{Cohen2000}.

% Nevertheless, in a limited space, the process of squeezing continues to the extent so that a saturation regime~\cite{Forgerini2014}, or characteristic time~\cite{pone} emerges. Thus, the \textit{duopoly} contest encounter with a zero-sum game, that is, customers as scarce resources distribute among the firms, and it may cause some advertising~\cite{Chintagunta1994,Ardalankia2019}, and competition costs. The more scarcity of the customers, the higher the possibility of the zero-sum game.

% When it comes to the role of strategy in a contest, it stems from some different internal and external aspects~\cite{Porter1989,Naik2005}. As proof of this concept, the underdogs may be united--in a cooperative process--to overcome the stronger side~\cite{Iranzo2016}. Besides, for the sake of achieving a winning position in the market share, marketing strategies may be changed by the managers in some occasions. In this perspective, the firms upon their ability to invest and their internal and environmental situations apply defensive or offensive marketing strategies to boost their market share~\cite{MartnHerrn2012,Bridges2009,Woodall2004,Erickson1993}.

We have utilized the same memory, that is, the same fractional derivative order, for both starting points--the initial time and the peak. Nonetheless, for further interpretation, it would be interesting to expand the meaning of growth rates and the concept of memory (or the fractional derivative order) of the proposed model in different contexts. For more realistic modelings, we can exploit the selective recalling-forgetting strategy with variable fractional order $\alpha(t)$ for a different position, rather than the peak point. 
%%%%%%%%%%%%%%%%%%%%%%%%%%%%%%%%%%%%%%%%%%%%%%%%%%%%%%%%%%%%%%%%%%%%%
\bibliography{editedbib.bib}

\begin{thebibliography}{10}

\bibitem{Saeedian2017}
M.~Saeedian, M.~Khalighi, N.~Azimi-Tafreshi, G.~R. Jafari, and M.~Ausloos,
  ``Memory effects on epidemic evolution: The susceptible-infected-recovered
  epidemic model,'' {\em
  \href{https://doi.org/10.1103/physreve.95.022409}{Physical Review E}},
  vol.~95, Feb. 2017.

\bibitem{banerjee2014gossip}
A.~Banerjee, A.~G. Chandrasekhar, E.~Duflo, and M.~O. Jackson, ``Gossip:
  Identifying central individuals in a social network,'' tech. rep., National
  Bureau of Economic Research, 2014.

\bibitem{bomze1995lotka}
I.~M. Bomze, ``Lotka-volterra equation and replicator dynamics: new issues in
  classification,'' {\em Biological cybernetics}, vol.~72, no.~5, pp.~447--453,
  1995.

\bibitem{gonccalves2013analytical}
G.~A. Gon{\c{c}}alves, R.~S. de~Quadros, and D.~Buske, ``An analytical
  formulation for pollutant dispersion simulation in the atmospheric boundary
  layer,'' {\em Journal of Environmental Protection}, vol.~4, no.~08, p.~57,
  2013.

\bibitem{cussler2009diffusion}
E.~L. Cussler and E.~L. Cussler, {\em Diffusion: mass transfer in fluid
  systems}.
\newblock Cambridge university press, 2009.

\bibitem{Ebadi2016}
H.~Ebadi, M.~Saeedian, M.~Ausloos, and G.~R. Jafari, ``Effect of memory in
  non-markovian boolean networks illustrated with a case study: A cell cycling
  process,'' {\em \href{https://doi.org/10.1209/0295-5075/116/30004}{{EPL}
  (Europhysics Letters)}}, vol.~116, p.~30004, Nov. 2016.

\bibitem{pone}
H.~Safdari, M.~Zare~Kamali, A.~Shirazi, M.~Khalighi, G.~Jafari, and M.~Ausloos,
  ``Fractional dynamics of network growth constrained by aging node
  interactions,'' {\em
  \href{https://journals.plos.org/plosone/article?id=10.1371/journal.pone.0154983}{PLOS
  One}}, vol.~11, pp.~1--13, 05 2016.

\bibitem{Iranzo2016}
J.~Iranzo, J.~M. Buld{\'{u}}, and J.~Aguirre, ``Competition among networks
  highlights the power of the weak,'' {\em
  \href{https://doi.org/10.1038/ncomms13273}{Nature Communications}}, vol.~7,
  Nov 2016.

\bibitem{Iori2015}
G.~Iori, R.~N. Mantegna, L.~Marotta, S.~Miccich{\`{e}}, J.~Porter, and
  M.~Tumminello, ``Networked relationships in the e-{MID} interbank market: A
  trading model with memory,'' {\em
  \href{https://doi.org/10.1016/j.jedc.2014.08.016}{Journal of Economic
  Dynamics and Control}}, vol.~50, pp.~98--116, jan 2015.

\bibitem{Thackeray2008}
R.~Thackeray, B.~L. Neiger, C.~L. Hanson, and J.~F. McKenzie, ``Enhancing
  promotional strategies within social marketing programs: Use of web 2.0
  social media,'' {\em \href{https://doi.org/10.1177/1524839908325335}{Health
  Promotion Practice}}, vol.~9, pp.~338--343, Mar. 2008.

\bibitem{Ashley2014}
C.~Ashley and T.~Tuten, ``Creative strategies in social media marketing: An
  exploratory study of branded social content and consumer engagement,'' {\em
  \href {https://doi.org/10.1002/mar.20761}{Psychology {\&} Marketing}},
  vol.~32, pp.~15--27, Dec. 2014.

\bibitem{Kaushik2019}
P.~Kaushik and N.~Kukreja, ``Study on recent trends in sales promotion in
  india,'' {\em
  \href{http://www.mbajournals.in/index.php/JoITM/article/view/243}{NOLEGEIN-
  Journal of Information Technology and Management}}, pp.~26--30, 2019.

\bibitem{Kaplan2013}
K.~Kaplan, ``Crowd-funding: Cash on demand,'' {\em
  \href{https://doi.org/10.1038/nj7447-147a}{Nature}}, vol.~497, pp.~147--149,
  May 2013.

\bibitem{podlubny1999fractional}
I.~Podlubny, ``Fractional differential equations, acad,'' {\em Press, London},
  p.~E2, 1999.

\bibitem{kilbas2006theory}
A.~A.~A. Kilbas, H.~M. Srivastava, and J.~J. Trujillo, {\em Theory and
  applications of fractional differential equations}, vol.~204.
\newblock Elsevier Science Limited, 2006.

\bibitem{Hassanibesheli2017}
F.~Hassanibesheli, L.~Hedayatifar, H.~Safdari, M.~Ausloos, and G.~Jafari,
  ``Glassy states of aging social networks,'' {\em Entropy}, vol.~19, p.~246,
  May 2017.

\bibitem{samko1993fractional}
S.~G. Samko, A.~A. Kilbas, O.~I. Marichev, {\em et~al.}, {\em Fractional
  integrals and derivatives}, vol.~1993.
\newblock Gordon and Breach Science Publishers, Yverdon Yverdon-les-Bains,
  Switzerland, 1993.

\bibitem{fabrizio1}
M.~Caputo and M.~Fabrizio, ``A new definition of fractional derivative without
  singular kernel,'' {\em
  \href{https://www.semanticscholar.org/paper/A-new-Definition-of-Fractional-Derivative-without-Caputo-Fabrizio/d658598a993790772bfae1d5a1c5fe33aefb773c}{Progr.
  Fract. Differ. Appl}}, vol.~1, no.~2, pp.~1--13, 2015.

\bibitem{atangana2016chaos}
A.~Atangana and I.~Koca, ``Chaos in a simple nonlinear system with
  atangana--baleanu derivatives with fractional order,'' {\em Chaos, Solitons
  {\&} Fractals}, vol.~89, pp.~447--454, 2016.

\bibitem{Krishnamoorthy2010}
A.~Krishnamoorthy, A.~Prasad, and S.~P. Sethi, ``Optimal pricing and
  advertising in a durable-good duopoly,'' {\em \href
  {https://doi.org/10.1016/j.ejor.2009.01.003}{European Journal of Operational
  Research}}, vol.~200, pp.~486--497, Jan. 2010.

\bibitem{Hu2010}
J.-L. Hu and C.-Y. Fang, ``Do market share and efficiency matter for each
  other? an application of the zero-sum gains data envelopment analysis,'' {\em
  \href {https://doi.org/10.1057/jors.2009.11}{Journal of the Operational
  Research Society}}, vol.~61, pp.~647--657, Apr. 2010.

\bibitem{Ardalankia2019}
J.~Ardalankia, M.~Osoolian, E.~Haven, and G.~Jafari, ``Multiscale features of
  cross correlation of price and trading volume,'' {\em Arxiv}, no.~1903.01744,
  2019.

\bibitem{Grauwin2009}
S.~Grauwin, E.~Bertin, R.~Lemoy, and P.~Jensen, ``Competition between
  collective and individual dynamics,'' {\em
  \href{https://doi.org/10.1073/pnas.0906263106}{Proceedings of the National
  Academy of Sciences}}, vol.~106, pp.~20622--20626, nov 2009.

\bibitem{Banerjee2013}
A.~Banerjee, A.~G. Chandrasekhar, E.~Duflo, and M.~O. Jackson, ``The diffusion
  of microfinance,'' {\em
  \href{https://doi.org/10.1126/science.1236498}{Science}}, vol.~341,
  pp.~1236498--1236498, July 2013.

\bibitem{Newman2001}
M.~E.~J. Newman, ``Clustering and preferential attachment in growing
  networks,'' {\em \href{https://doi.org/10.1103/physreve.64.025102}{Physical
  Review E}}, vol.~64, jul 2001.

\bibitem{Newman2003}
M.~E.~J. Newman, ``The structure and function of complex networks,'' {\em
  \href{https://doi.org/10.1137/s003614450342480}{{SIAM} Review}}, vol.~45,
  pp.~167--256, jan 2003.

\bibitem{Barabsi1999}
A.-L. Barab{\'{a}}si, R.~Albert, and H.~Jeong, ``Mean-field theory for
  scale-free random networks,'' {\em
  \href{https://doi.org/10.1016/s0378-4371(99)00291-5}{Physica A: Statistical
  Mechanics and its Applications}}, vol.~272, pp.~173--187, oct 1999.

\bibitem{Barabaacutesi1999}
A.-L. Barab{\'{a}}si and R.~Albert, ``Emergence of scaling in random
  networks,'' {\em
  \href{https://doi.org/10.1126/science.286.5439.509}{Science}}, vol.~286,
  pp.~509--512, oct 1999.

\bibitem{Forgerini2014}
F.~L. Forgerini and N.~Crokidakis, ``Competition and evolution in restricted
  space,'' {\em \href{https://doi.org/10.1088/1742-5468/2014/07/p07016}{Journal
  of Statistical Mechanics: Theory and Experiment}}, vol.~2014, p.~P07016, jul
  2014.

\bibitem{Cohen2000}
R.~Cohen, K.~Erez, D.~ben Avraham, and S.~Havlin, ``Resilience of the internet
  to random breakdowns,'' {\em \href
  {https://doi.org/10.1103/physrevlett.85.4626}{Physical Review Letters}},
  vol.~85, pp.~4626--4628, nov 2000.

\bibitem{Diethelm2004}
K.~Diethelm, N.~J. Ford, and A.~D. Freed, ``Detailed error analysis for a
  fractional adams method,'' {\em Numerical Algorithms}, vol.~36, pp.~31--52,
  May 2004.

\bibitem{Garrappa}
R.~Garrappa, ``On linear stability of predictor–corrector algorithms for
  fractional differential equations,'' {\em International Journal of Computer
  Mathematics}, vol.~87, no.~10, pp.~2281--2290, 2010.

\end{thebibliography}
\bibliographystyle{ieeetr}

\section*{Appendix}
{\centering\textbf{Numerical solution}}\\

Incommensurate fractional differential equations \ref{eq:11}-\ref{eq:4} can be written as:
\begin{equation}\label{eq:5}
{}_{{t_0}}^cD^{\bf{a}}_{t}{\bf{y}}(t)={\bf{f}}(t,{\bf{y}}(t)).
\end{equation}

The vectors ${\bf{y}}=(y^1,y^2,y^3)$ and ${\bf{f}}=(f_1,f_2,f_3)$ are corresponding to $(S,I_1,I_2)$ and their function of differentials, respectively, and ${\bf{a}}=(\alpha_1,\alpha_2,\alpha_3)$ denotes the orders of the differential equations such that $\alpha_1=\alpha_2=1$. Notice that the system \ref{eq:11}-\ref{eq:4} is a generalized form of the system \ref{eq:1}-\ref{eq:3}. Therefore, the solution of the latter system is a particular solution of the former one, when $\alpha_3=1$. The numerical solution of such equations comes from the discretization of an equivalent Volterra integral equation which is extensively presented in~\cite{Diethelm2004,Garrappa}:

\begin{equation}\label{eq:6}
y^{i}_{n}=y^{i}_{0}+h^{\alpha_{i}}\Sigma^{n-1}_{k=0}b_{n-k-1}^{i}f^{i}_{k}.
\end{equation}

In the numerical solution, the time is discretized as $T={t_0,\dots,t_n}$ where $t_n=hn$ and $h$ is the step size. The recursive Eq.\ref{eq:6} gives the value of $y^{i}$ at time $n$ based on the initial states $y^{i}_0$ and the solutions of the Eq.\ref{eq:5} at the prior time steps of functions $f^{i}_{k}$ with weight $b_{n-k-1}^{i}$. Hence, an explicit scheme gives the weight coefficient as follow~\cite{Diethelm2004}:

\begin{equation}\label{eq:7}
b_{n-k-1}^{i}= \frac{(n-1-k)^{\alpha_i}-(n-k)^{\alpha_i}}{\Gamma(\alpha_i+1)}.
\end{equation}

\end{document}